\documentclass[final,numberedheadings]{aipproc}

\layoutstyle{8x11single}

\input{epsf}

\def\CQG{{\it Class. Quantum Gravity} }

\def\GRG{{\it Gen. Relativity and Gravitation} }

\def\JHEP{{\it JHEP} }

\def\MPL{{\it Mod. Phys. Lett.} }

\def\PL{{\it Phys. Lett.} }
\def\PR{{\it Phys. Rev.} }
\def\PRL{{\it Phys. Rev. Lett.} }

\def\PTP{{\it Progr. Theor. Phys.} }

\def\al{\alpha}
\def\be{\beta}
\def\ga{\gamma}
\def\de{\delta}

\def\th{\theta}

\def\ka{\kappa}
\def\la{\lambda}

\def\Ga{\Gamma}
\def\De{\Delta}

\def\mn{{\mu\nu}}

 \def\frac#1#2{{\textstyle{{#1}\over
{#2}}}} 
\def\lsim{\mathrel{\rlap{\lower4pt\hbox{\hskip1pt$\sim$}}
    \raise1pt\hbox{$<$}}} \def\gsim{\mathrel{\rlap{\lower4pt\hbox{\hskip1pt$\sim$}}
    \raise1pt\hbox{$>$}}}
\def\sqr#1#2{{\vcenter{\vbox{\hrule height.#2pt
         \hbox{\vrule width.#2pt height#1pt \kern#1pt
         \vrule width.#2pt}
         \hrule height.#2pt}}}} 

\def\beq{\begin{equation}}
\def\eeq{\end{equation}}
\def\beqa{\begin{eqnarray}}
\def\eeqa{\end{eqnarray}}

\begin{document}

\title{The flight of the bumblebee: solutions from a vector-induced
spontaneous Lorentz symmetry breaking model}



\author{Orfeu Bertolami}{
  address={Instituto Superior T\'ecnico, Departamento de
F\'{\i}sica, \\ Av. Rovisco Pais 1, 1049-001 Lisboa, Portugal} }

\author{Jorge P\'aramos}{
  address={Instituto Superior T\'ecnico, Departamento de
F\'{\i}sica, \\ Av. Rovisco Pais 1, 1049-001 Lisboa, Portugal} }

\begin{abstract}
The vacuum solutions arising from a spontaneous breaking of
Lorentz symmetry due to the acquisition of a vacuum expectation
value by a vector field are derived. These include the purely
radial Lorentz symmetry breaking (LSB), radial/temporal LSB and
axial/temporal LSB scenarios. It is found that the purely radial
LSB case gives rise to new black hole solutions. Whenever
possible, Parametrized Post-Newtonian (PPN) parameters are
computed and compared to observational bounds, in order to
constrain the Lorentz symmetry breaking scale.
\end{abstract}

\maketitle


\section{Introduction}

Lorentz invariance is clearly one of the most fundamental
symmetries of Nature. It is both theoretically sound and
experimentally well tested \cite{Kostelecky1, Bertolami}, thus
playing a leading role in most theories of gravity. Therefore, it
is only natural that little attention has been paid to the
consequences of explicitly breaking this symmetry.

A more flexible approach to this question admits a spontaneous
breaking of this symmetry, instead of an explicit one, analogously
to the Higgs mechanism in the Standard Model of particle physics
\cite{Kostelecky2,Kostelecky3,Kostelecky4, Kostelecky5}. This can
arise if a vector field ruled by a potential exhibiting a minimum
rolls to its vacuum expectation value (\textit{vev}) -- this
vector field, usually referred to as ``bumblebee'' vector, thus
acquires a specific four-dimensional orientation.

From a theoretical standpoint, a spontaneous Lorentz symmetry
breaking (LSB) is possible, for instance, in string/M-theory,
arising from non-trivial solutions in string field theory
\cite{Kostelecky2, Kostelecky3} and in noncommutative field
theories \cite{LSB4, LSB5}. A spacetime variation of fundamental
coupling constants could also lead to a spontaneous LSB
\cite{LSB6}. Experimentally, the violation of Lorentz invariance
could be tested in ultra-high energy cosmic rays \cite{LSB7}.

The consequences of the ``bumblebee'' vector scenario were studied
in Ref. \cite{paper}; in there, three relevant cases were taken
into account: the bumblebee field acquiring a purely radial
\textit{vev}, a mixed radial and temporal \textit{vev} and a mixed
axial and temporal \textit{vev}. The results were analyzed in
terms of the PPN parameters, when possible, prompting for
comparison with current and future experimental bounds and
effects, for instance, from string theory in a low-energy
scenario. These bounds may arise from the observations of the
Bepi-Colombo \cite{Bepi} and LATOR \cite{LATOR} missions (see also
Ref. {\cite{paper2}) for a discussion on future gravitational
experiments).

The action of the bumblebee model is written as

\beq S = \int d^4 x \sqrt{-g} \left[{1 \over 2 \ka} \left( R + \xi
B^\mu B^\nu R_{\mu\nu} \right) - {1 \over 4} B^{\mu\nu} B_{\mu\nu} -
V(B^\mu B_\mu \pm b^2 ) \right]~~, \eeq

\noindent where $\ka = 8 \pi G$, $B_{\mu\nu} = \partial_\mu B_\nu
- \partial_\nu B_\mu$, $\xi$ is a coupling constant and $b^2$ sets
the bumblebee's \textit{vev}, since the potential $V$ driving
Lorentz and/or CPT violation is supposed to have a minimum at $
B^\mu B_\mu \pm b^2 = 0 $, \textit{i.e.} $V'(b_\mu b^\mu)=0$. The
particular form of this potential is irrelevant, since one assumes
that the bumblebee field is frozen at its \textit{vev}. The scale
of $b_\mu$ should be obtained from string theory or from a
low-energy extension to the Standard Model. Hence, one expects
$b_\mu$ to be of order of $M_{Pl}$, the Planck mass, or $M_{EW}$,
the electroweak breaking scale.

\section{Purely radial LSB}

In this section, a method to obtain the exact solution for the
purely radial LSB is developed. A static, spherically symmetric
spacetime, with a Birkhoff metric $ g_\mn = diag(-e^{2\phi},e^{2
\rho},r^2,r^2 sin^2\th)$ is considered. It can be easily seen that
the Killing vectors of the metric are conserved, showing that
radial symmetry is still valid; this enables the construction of a
covariantly conserved current associated with the vector bumblebee
field \cite{paper}.

The affine connection derived from the metric $g_\mn$ allows for
the computation of $b_\mu$, given the non-trivial covariant
derivative with respect to the radial coordinate, and taking
$b_\mu = (0, b(r), 0, 0)$. Hence, from $D_\mu b_\nu =
\partial_\mu b_\nu - \Ga^\al_{\nu\mu} b_\al = 0$, it follows that
$ b(r) = \xi^{-1/2} b_0 e^{\rho}$, where the factor $ \sqrt{\xi}$
is introduced for later convenience. As expected, $ b^2 = b^\mu
b_\mu = b_0^2 \xi^{-1} $ is constant.

The (spatial) action can be thus written as

\beq S_s = \int d^4 x \sqrt{-g} \left[ {R \over 2 \ka} +
(g^{rr})^2 b^2(r) R_{rr} \right]_s ~~. \eeq

\noindent The determinant of the metric is given by $\sqrt{-g} =
r^2 e^{\rho + \phi}$; the scalar curvature and the relevant
non-vanishing Ricci tensor component are given by $ R = 2 r^{-2}
\left[1 + (2 r \rho' -1) e^{-2\rho} \right]$ and $R_{rr} = 2
r^{-1} \rho'$, where the prime stands for derivative with respect
to $r$ and we have integrated over the angular dependence. Also, $
\xi (b^r)^2 R_{rr} = b_0^2 2 r^{-1} \rho' e^{-2\rho}$, where $b^r$
is the contravariant radial component of $b_\mu$. By introducing
the field redefinition $ \Psi = \left(1-e^{-2\rho} \right)
r^{-2}$, the action may be rewritten as \cite{Bento}

\beq S_s = {2 \over \ka} \int dr ~e^{\rho + \phi} r^2 \left[ ( 3 +
b_0^2 ) \Psi + ( 1 + {b_0^2 \over 2} ) r \Psi' \right]~~.
\label{spaction} \eeq

Variation with respect to $\phi$ produces the equation of motion

\beq ( 3 + b_0^2 ) \Psi + ( + 1 + {b_0^2 \over 2} ) r \Psi' = 0
~~, \eeq

\noindent which admits the solution $ \Psi(r) = \Psi_0 r^{-3+L}$,
with $ L \equiv 3 - (3 + b_0^2) / ( 1 + b_0^2 / 2) \simeq b_0^2 /
2$, and hence $L \simeq b_0^2 /2$. We thus obtain $ g_{rr} = e^{2
\rho} = \left( 1 - \Psi_0 r^{-1+L} \right)^{-1}$. Comparison with
the usual Schwarzschild metric yields

\beq g_{rr} = \left( 1 - {2 G_L m \over r} r^L \right)^{-1}~~,~~
g_{00} = -1 + {2 G_L m \over r} r^L~~, \label{metric} \eeq

\noindent where $G_L$ has dimensions $ [ G_L ] = L^{2-L} $ (in
natural units, where $c=\hbar=1$); one can define $G_L = G
r_0^{-L}$, where $r_0$ is an arbitrary distance. The $L
\rightarrow 0$ limit yields $G_L \rightarrow G$ and the usual
geometrical mass, $M \equiv G m $, with dimensions of lenght. From
now on we express all results in terms of $M$. The event horizon
condition is given by

\beq g_{00} = -1 + {2 M \over r} {r \over r_0}^L = 0 ~~, \eeq

\noindent thus $ r_s = (2 M r_0^{-L} )^{1/1-L} $. The norm-square
of the Riemann tensor, $I = R_{\mu \nu \rho \la}R^{\mu \nu \rho
\la}$, is found to be

\beq I = 48 \left[ 1 - {5 \over 3} L + {17 \over 12} L^2 - {1
\over 2} L^3 + {1 \over 12} L^4 \right] M^2 \left({r \over
r_0}\right)^{2L} r^{-6} \simeq \left( 1 - {5 \over 3} L\right)
\left({r \over r_0}\right)^{2 L} I_0 ~~, \eeq

\noindent where $I_0 = 48 M^2 r^{-6}$ is the usual scalar
invariant in the limit $L \rightarrow 0$. Since $I(r=r_s)$ is
finite, the singularity at $r = r_s$ is removable; accordingly,
the singularity at $r=0$ is intrinsic, as the scalar invariant
diverges there. One concludes that an axial LSB gravity model
admits new black hole solutions with a singularity well protected
within a horizon of radius $r_s$. The associated Hawking
temperature is

\beq T = {\hbar \over k_B} {1 \over 4 \pi r_s } = (2M
r_0^{-L})^{-L/(1-L)} T_0 \simeq (2M r_0^{-L})^{-L} T_0 ~~, \eeq

\noindent where $T_0 = \hbar / 8 \pi k_B M $ is the usual Hawking
temperature, recovered in the limit $L \rightarrow 0$.

Since the obtained metric cannot be expanded in powers of $U=M/r$,
a PPN expansion is not feasible. However, a comparison with
results for deviations from Newtonian gravity \cite{Fischbach},
usually stated in terms of a Yukawa potential of the form

\beq V_Y(r) = {G_Y m \over r} \left( 1 + \al e^{-r/\la} \right)~~,
\eeq

\noindent yields $G_L r^L = G_Y \left(1 + \al e^{-r/\la} \right)$,
which, to first order around $r = r_0$, reads

\beq G_L r_0^L \left(1 + L {r \over r_0} \right) = G_Y \left(1 +
\al - \al {r \over \la} \right)~~, \eeq

\noindent so that one identifies $\la = r_0$ and $\al = -L$ (with
$G_Y (1-L) = G_L r_0^L = G$). Planetary tests to Kepler's law in
Venus indicate that $\la = r_0 = 0.723~AU$ and $L = |\al| \leq 2
\times 10^{-9}$.

\section{Radial/temporal LSB}

We consider now the mixed radial and temporal Lorentz symmetry
breaking. As before, it is assumed that the bumblebee field
$B_\mu$ has relaxed to its vacuum expectation value. Provided that
one takes temporal variations to be of the order of the age of the
Universe $H_0^{-1}$, where $H_0$ is the Hubble constant, a
Birkhoff static, radially symmetric metric $ g_\mn =
diag(-e^{2\phi},e^{2 \rho},r^2,r^2 sin^2(\th))$ may still be used.
The physical gauge choice of a vanishing covariant derivative of
the field $B_\mu$ yields $b_r(r) = \xi^{-1/2} A_r e^{\rho}$ and,
similarly, $ b_0(r) = \xi^{-1/2} A_0 e^{\phi}$, with $A_0$ and
$A_r$ dimensionless constants. As before, $ b^2 = b^\mu b_\mu =
(A_r^2 - A_0^2)\xi^{-1}$ is constant.

In the present case, the symmetry $\phi = -\rho$ does not hold, as
now both a radial and a temporal component for the vector field
\textit{vev} are present; for this reason, the previous spatial
action formalism depicted in Eq. (\ref{spaction}) cannot be used.
Instead, the full Einstein equations must be dealt with,

\beq G_{\mu\nu} = \xi \left[ {1 \over 2} (b^\al)^2 R_{\al\al}
g_{\mu\nu} - b_\mu b^\nu R_{\mu\nu} - b_\nu b^\mu R_{\nu\mu}
\right] ~~.\eeq

Since the bumblebee field has relaxed to its \textit{vev} and
therefore both the field strength and the potential term vanish,
the additional equation of motion for the vector field is trivial.
The metric Ansatz and the expressions for $b_\mu$ then yield

\beq G_{00} = {1\over 2} \left[3 A_0^2 R_{00} - A_r^2 e^{2(\phi
-\rho)} R_{rr} \right] ~~,~~ G_{rr} = {1\over 2} \left[A_0^2 e^{2
(\rho -\phi)} R_{00} - 3 A_r^2 R_{rr} \right] ~~,\eeq

\noindent Writing $G_{00} = g_0(r) e^{2(\phi-\rho)}$, $G_{rr} =
g_r(r)$, $R_{00} = f_0(r) e^{2(\phi -\rho)}$ and $R_{rr} =
f_r(r)$, where

\beqa \nonumber f_0(r) & \equiv & {(2 - r \rho')\phi' \over r} + \phi'^2 + \phi''~~,
~~ f_r(r) \equiv {(2+r\phi')\rho' \over r} - \phi'^2 - \phi''~~, \\
\nonumber g_0(r) & \equiv & {-1 + e^{2 \rho} \over r^2} + {2\rho'
\over r}~~,~~ g_r(r) \equiv {1 - e^{2 \rho} \over r^2} + {2\phi'
\over r}~~. \eeqa

\noindent the Einstein equations read

\beq \label{coupled} g_0(r) = {1\over 2} \left[3 A_0^2 f_0(r) -
A_r^2 f_r(r) \right]~~, ~~ g_r(r) = {1\over 2} \left[A_0^2 f_0(r)
- 3 A_r^2 f_r(r) \right]~~. \eeq

\noindent This is the set of coupled second order differential
equations which must be solved, with boundary conditions given by
$\phi(\infty) = \rho(\infty) = \phi'(\infty) = \rho'(\infty) = 0$.

The spontaneous LSB is clearly exhibited; as can be noticed from
$g_0 + g_r = f_0 + f_r $, one has $ (1 - 2A_0^2) f_0 = - ( 1 +
2A_r^2) f_r$; in the unperturbed case $A_0 = A_r = 0$, $f_0 = -
f_r$, and the Schwarzschild solution $\phi = -\rho$ is recovered
from $g_0 + g_r = 0 $. This symmetry does not hold in the
perturbed case, which produces $f_0 \approx
-(1+2A_0^2+2A_r^2)f_r$.

An expansion of the metric in terms of $\phi=\phi_0 + \de \phi$
and $\rho = - \phi_0 - \de \rho$ allows for the solving of Eqs.
(\ref{coupled}), where $\phi_0$ is given by the usual
Szcharzschild metric, $ \phi_0 = {1 \over 2} ln \left(1- 2M / r
\right)$, and $\de \rho$, $\de \phi$ are assumed to be small
perturbations. After some algebra, the solution is found to be
\cite{paper}

\beq \de \phi = K r^{-\al}~~, ~~ \de \rho \simeq \left[ 1 +
\al{(A+B) \over 2} \right] K r^{-\al} ~~, \eeq

\noindent where $A=A_0^2$, $B=A_r^2$, $K$ is an integration
constant and

\beq \al = {-C_1 - C_2 + \sqrt{(C_1 + C_2)^2 + 4 C_1 C_3} \over 2
C_1} > 0~~, \eeq

\noindent with

\beq \nonumber C_1 = A + 3B +AB + 9 B^2 \simeq A + 3B ~~, ~~ C_2=
2+B-3A+16AB \simeq 2~~,~~ C_3 = 2+B-7A \simeq 2 ~~. \eeq

\noindent One can linearize the exponent $\al$ around $C_1 \ll 1$,
yielding $ \al \simeq (2-7A+B)/(2+A+3B)$, so that $\al \simeq 1$.

After solving the coupled differential Eqs. (\ref{coupled}), the
non-trivial components of the metric now read

\beqa g_{tt} &=& -e^{2 (\phi_0 + \de \phi)} = -e^{2K r^{-\al}}
\left( 1 - {2M \over r} \right)~~, \\ \nonumber g_{rr} &=&
e^{-2(\phi_0 + \de \rho)} = {e^{-\left(2 + \al(A+B) \right) K
r^{-\al}} \over 1 - {2M \over r}} \equiv {e^{-2K_r r^{-\al}} \over
1 - {2M \over r}} ~~, \label{smetric} \eeqa

\noindent with the definition $K_r \equiv [1 + \al(A+B)/2]K \simeq
[1 + (A+B)/2]K$. Following the algebra of a Lorentz transformation
to a isotropic coordinate system, on which all spatial metric
components are equal, and then to a quasi-cartesian referential,
the resulting metric is \cite{paper}

\beq \eta_{tt} = g_{tt} = -1 + 2 U - \left(1- {K + K_r \over M}
\right) U^2~~,~~ \eta_{\xi' \xi'} = 1 + \left(1 - {K + 2 K_r \over
M} \right) U ~~, \eeq

\noindent and the PPN parameters may be directly read, yielding $
\be \simeq 1 - (K + K_r)/M$ and $\ga \simeq 1 - (K + 2 K_r)/M$.
Inverting this relation gives $ K / M \simeq 1 - 2 \be + \ga$,
$K_r / M \simeq \be - \ga$. Hence, a temporal/radial LSB manifests
itself linearly on the PPN parameters $\be$ and $\ga$. A caveat of
these results is the clear dependence of the obtained PPN
parameters on the free-valued integration constants $K$ and $K_r$,
instead of the physical parameters associated with the breaking of
Lorentz invariance. This reflects the linearization procedure
followed in order to obtain the radially symmetric Birkhoff metric
solution to the Einstein equations.

The bounds derived from the Nordvedt effect, $| \be -1 | \leq 6
\times 10^{-4}$ \cite{Will2} and the Cassini-Huygens experiment, $
\ga = 1 + (2.1 \pm 2.3) \times 10^{-5}$ \cite{Bertotti}, can be
used to obtain $ |K + K_r| < 0.9~m$ and $ K + 2 K_r = (-3.1 \pm
3.4) \times 10^{-2}~m$. Since, by definition

\beq K_r = \left[ 1 + \al{(A+B) \over 2} \right] K~~, \eeq

\noindent with $\al \simeq 1$, $A , B \ll 1$, deviations of $K_r$
from $K$ are expected to be small. Thus, considering for instance
the constraint $|1 - K_r / K | < 0.1$, one gets $\al(A+B) < 0.2$;
the limiting case $K \sim K_r$ gives $K \simeq (-1 \pm 1.1) \times
10^{-2}$, indicating a perturbation with a very short range
(actually, well inside the Sun, so that one should work with the
interior Scharzschild solution). The range of allowed values for
these parameters is depicted in Figures $1$ and $2$.


\begin{figure}
  \includegraphics[height=6cm]{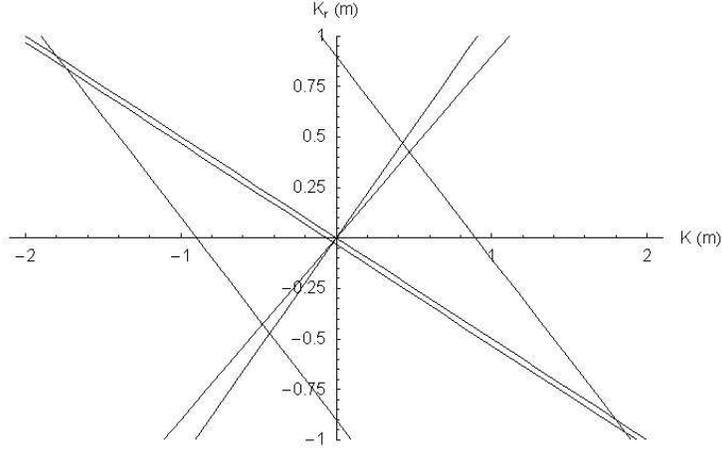}
  \caption{Allowed values for $K$ and $K_r$ \cite{paper}.}
\end{figure}



\begin{figure}
  \includegraphics[height=6cm]{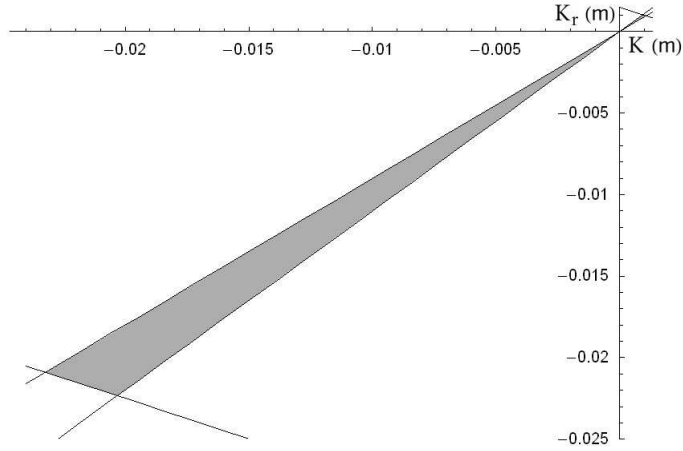}
  \caption{Detail of Fig. $1$, showing only the allowed
region \cite{paper}.}
\end{figure}


In the limit $M \rightarrow 0$, Eqs. (\ref{smetric}) yield
$\bar{g}_{tt} = -e^{2K / \xi}$ and $\bar{g}_{rr} = -e^{2K_r /
\xi}$. An analogy with Rosen's bimetric theory allows for the PPN
parameter $\al_2$ to be obtained, by interpreting this change of
the metric component as due to a background metric $\eta_{\mu\nu}$
\cite{Rosen}. Notice, however, that the vector field no longer
rolls to a radial \textit{vev} in the absence of a central mass,
since this spatial symmetry is inherited from its presence, so
this result should be taken with caution; this said, one obtains
\cite{Will} $ \al_2 = (c / c_G) -1 = e^{2(K_r - K) / \xi} - 1
\simeq 2 (K_r - K) / \xi = (A + B) \al / \xi$, which has a radial
dependency. Assuming $\al \simeq 1$ and considering the spin
precession constraint arising from solar to ecliptic alignment
measurements \cite{Will2}, one has $| \al_2 | = |A+B|/r < 4 \times
10^{-7}$, implying that $|A+B| < 4 \times 10^{-7} r_\odot = 2.78
\times 10^2~m $, where $r_\odot = 6.96 \times 10^{8}~m$ is the
radius of the Sun.

Further pursuing this analogy, we remark that, since there is no
explicit Lorentz breaking, the speed of light remains equal to
$c$. However, the speed of gravitational waves $c_G$ is shifted by
an amount

\beq \sqrt{c_G \over c} = e^{2(K-K_r) / \xi} \simeq \left[ 1 - {A
+ B \over \xi} \right]~~,\eeq

\noindent and hence it acquires a radial dependence. As stated
before, this result is highly simplistic and should be taken with
caution, since it lacks a complete treatment of gravitational
radiation induced by LSB, taking into account variations of the
bumblebee field $B_\mu$ around its \textit{vev} $b_\mu$.

Finally, notice that, since the radial LSB effects are exact,
while the radial/temporal results are not, a direct comparison of
these scenarios by taking the $A \rightarrow 0$ limit is not
possible.

\section{Axial/temporal LSB}

The anisotropic LSB case is dealt with in this section. As before,
we assume that the bumblebee field is stabilized at its vacuum
expectation value, which possesses both a temporal and a spatial
component; the latter is taken to lie on the x-axis, that is, $
b_\mu = \ka^{-1} (a,b,0,0) $. Since the radial symmetry of the
Scharzschild is clearly broken, one cannot resort to a Birkhoff
canonical \textit{Ansatz}. Instead, the perturbations $h_\mn$ to
the flat Minkowsky metric must be obtained. To first order in
$h_\mn$, one has

\beqa && R_{00} = - {1 \over 2} \nabla^2 h_{00}~~, ~~ R_{0i} = -{1
\over 2} \left( \nabla^2 h_{0i} - h_{k0,ik} \right) ~~,
\\ \nonumber && R_{ij} = -{1 \over 2} \left( \nabla ^2 h_{ij} - h_{00,ij} +
h_{kk,ij} - h_{ki,kj} - h_{kj,ki} \right) ~~, \eeqa

\noindent where time derivatives were neglected, since one assumes
that $v \ll c$.

In order to solve the Einstein equations, one first writes the
stress-energy tensor for the bumblebee field,

\beq T_{B \mn} = \left[ {1 \over 2} b^\al b^\be R_{\al \be} g_\mn
- b_\mu b^\al R_{\al \nu} - b_\nu b^\al R_{\al \mu} \right]~~,\eeq

\noindent which has a vanishing trace. From the trace of the
Einstein equations, one gets

\beq R_\mn = \ka \left[T_\mn + {1 \over 2} g_\mn T + T_{B \mn}
\right] ~~. \eeq

\noindent To get the $h_{00}$ component to first order in the
potential $U$, one writes

\beq R_{00} =  -{1 \over 2} \left( a^2 R_{00} + b^2 R_{11} + 2 a b
R_{10} \right) - 2 a \left( a R_{00} + b R_{10} \right) ~~, \eeq

\noindent which, after a little algebra \cite{paper}, yields the
differential equation

\beq \left({2 + 5 a^2 - b^2 \over 2 + 5 a^2} \right) h_{00,11} +
h_{00,22} + h_{00,33} = 0 ~~. \eeq

\noindent This admits the solution

\beq h_{00} (x,y,z) = {2 M \over \sqrt{c_0^2 x^2 + y^2 + z^2}} ~~,
\eeq

\noindent where $c_0^2 = ( 2 + 5 a^2) / (2 +5 a^2 - b^2)$.

Similarly, the $h_{ii}$ components ($i \neq 1$) obey

\beq \left( 2 - b^2 \right) h_{ii,11} + 4 h_{ii,ii} + 2 h_{ii,jj}
= \left( a^2 - b^2 \right) h_{00,11} + \left( 2 + a^2 \right)
h_{00,ii} + a^2 h_{00,jj} ~~. \label{partial} \eeq

\noindent Taking the Ansatz $h_{ii} (x,y,z) = h_{00} (\al_1 x,
\al_2 y, \al_3 z)$ and, after some calculation (see Appendix I of
Ref. \cite{paper}), one can obtain for the coefficients $\al_i$:

\beq \al_1^2 = 1 - {(2 - a^2) b^2 \over (2 + 5 a^2) ( 2 - b^2 )
}~~, ~~ \al_i = \al_j = 1 ~~. \eeq

\noindent Hence,

\beq h_{ii} (x,y,z) = h_{00} (\al_1 x, y, z) \equiv {2 M \over
\sqrt{c_2^2 x^2 + y^2 + z^2}} ~~, \eeq

\noindent with the definition

\beq c_2^2 = \al_1^2 c_0^2 = {2 (2 + 5a^2 - 2 b^2) - 4 a^2b^2
\over (2 + 5a^2 - b^2) (2 - b^2)} ~~. \eeq

The $h_{11}$ component is now computed; a similar calculation
leads to the differential equation \cite{paper}

\beq h_{11,22} + h_{11,33} = 2M \left[ \left({ a^2 (c_0^2 -1)
\over 2 + 3 b^2 } + c_0^2 \right) { 2 c_0^2 x^2 -y^2 -z^2 \over
(c_0^2 x^2 + y^2 + z^2 )^{5/2} }- 2 c_2^2 {2 c_2^2 x^2 -y^2 -z^2
\over (c_2^2 x^2 + y^2 + z^2 )^{5/2} } \right]~~, \eeq

\noindent indicating that the solution is a linear combination of
$h_{00}$ and $h_{22}$. Indeed,

\beq h_{11} (x,y,z) = - \left({ a^2 ( c_0^2 - 1 ) \over 2 + 3 b^2
} + c_0^2 \right) h_{00}(x,y,z) + 2 c_2^2 h_{22} (x,y,z)~~. \eeq

Proceeding to the off-diagonal component $h_{10}$, one obtains the
differential equation

\beq h_{01,22} + h_{01,33} = -{ab \over 1+ a^2 + b^2} \left[
h_{00,22} + h_{00,33} + h_{11,22} + h_{11,33} - 2 h_{22,11}
\right] ~~. \eeq

\noindent Writing $h_{01} = -ab/(1+a^2+b^2)(h_{00} + h_{11} + \de
h_{01})$ leads to

\beq \de h_{01,22} + \de h_{01,33} = 2 h_{22,11} = {4 M c_2^2 (2
c_2^2 x^2 - y^2 - z^2) \over (c_2^2 x^2 +y^2 +z^2)^{5/2}}~~, \eeq

\noindent and hence $ \de h_{01}(x,y,z) = - 2 c_2^2 h_{22}
(x,y,z)$. Therefore,

\beq h_{01} = {a b \over 1 + a^2 + b^2 } \left[{ a^2 ( c_0^2 - 1 )
\over 2 + 3 b^2 } + c_0^2 \right] h_{00} ~~ . \eeq

Finally, the $h_{00}$ component is computed to second order (see
Appendix II of Ref. \cite{paper}); it can be shown that only a
correction to the first order term $h_{00}^{(1)}$ appears:

\beq h_{00} = {2c_0^2 [6 + 9 b^2 + (15 + 22 b^2) a^2] + a^2 b^2
\over c_0^2 (6 + 15 a^2 + b^2) (2 +3b^2)} h_{00}^{(1)} \simeq
\left( 1 - {b^2 \over 6} \right) h_{00}^{(1)}~~.\eeq

The PPN formalism cannot be straightforwardly used to ascertain
its effects, since it relies on a transformation to a
quasi-cartesian frame of reference on which, by definition, all
diagonal metric components $g_{ii}$ are equal. However, some
PPN-like parameters may be extracted from the results, by noticing
that

\beq {1 \over \sqrt{(1+ 2 k) x^2 + y^2 + z^2}} \simeq {1 \over r}
\left( 1 - k cos^2 \th \right) ~~ . \eeq

For $h_{00}$, one gets

\beq h_{00} = \left( 1 - {b^2 \over 6} \right) 2M {1 - (c_0^2 - 1)
cos^2 \th \over r}~~. \label{anisotropy} \eeq

\noindent Since no $r^{-2}$ correction appears, the PPN parameter
$\be$ vanishes in this approach. However, as $h_{11} \neq h_{22} =
h_{33}$, the same reasoning allows two parameters analogous to the
$\ga$ PPN parameter to be obtained: after neglecting the
normalization with respect to $h_{00}$, one gets

\beqa \nonumber \ga_1 & = & 1 + cos^2 \th \times \left[ -\left[{
a^2 ( c_0^2 - 1 ) \over 2 + 3 b^2 } + c_0^2 \right] (1 - c_0^2) +
2 c_2^2 (1 -
c_2^2) \right] \simeq 1 + {b^2 \over 2} cos^2 \th ~~, \\
\ga_2 &=& 1 + (1 - c_2^2) cos^2 \th \simeq 1 - \left({a b \over 2}
\right)^2 cos^2 \th~~. \eeqa

\noindent As expected, the x-axis LSB produces a stronger effect
on the $h_{11}$ component. No clear connection can be derived to
link $\ga$ with $\ga_1$ and $\ga_2$, due to the aforementioned
anisotropy. However, one can take $\ga$ to be of the same order of
magnitude as the average of $\ga_1$ and $\ga_2$, integrated over
one orbit:

\beq \ga -1 \simeq {1 \over 2} (\ga_1 + \ga_2) - 1 \simeq  {b^2
\over 4} \langle cos^2 \th \rangle \simeq {b^2 \over 8}
(1-e^2)~~,\eeq

\noindent where $e$ is the orbit eccentricity. For low values of
$e$, one gets $\ga \simeq b^2 / 8$; The constraint $\ga =1 + (2.1
\pm 2.3) \times 10^{-5}$ then enables $|b| \leq 1.9 \times
10^{-2}$.

A discussion concerning the anisotropy of inertia and its effect
in the width of resonance lines has been presented as a test
between Mach's principle and the Equivalence Principle
\cite{Weinberg,Kostelecky6}, relying on the hypothetical effect on
the proton mass of the proximity to the galactic core. In the
present scenario, we note that a radial LSB with the galactic core
acting as source would amount to an axial LSB in a small region
such as the Solar System. The bound $\De m_P / m_P \leq 3 \times
10^{-22}$, $m_P$ being the proton mass \cite{Lamoreaux}, can then
be used to obtain

\beq {\De m_P \over m_P} = \left( 1 - {b^2 \over 6} \right) (c_0^2
-1 ) \simeq {b^2 \over 2} \leq 3 \times 10^{-22}~~,\eeq

\noindent resulting in the limit $ |b| \leq 2.4 \times 10^{-11} $,
a much more stringent bound than the obtained above.

\section{Conclusions}

In this contribution, the solutions of a gravity model coupled to
a vector field where Lorentz symmetry is spontaneously broken are
studied, and three different relevant scenarios were highlighted:
a purely radial, temporal/radial and temporal/axial LSB.

In the first case, a new black hole solution is found, exhibiting
a removable singularity at a horizon of radius $r_s = (2M
r_0^{-L})^{1/1-L}$, slightly perturbed with respect to the usual
Scharzschild radius $r_{s0} = 2M$. This has an associated Hawking
temperature of $T = (2M r_0^{-L})^{-L}T_0$, where $T_0 = \hbar/ 8
\pi \ka_B M$ is the usual Hawking temperature, and protects an
intrinsic singularity at $r=0$. Bounds on deviations from Kepler's
law yield $L \leq 2 \times 10^{-9}$.

The temporal/radial scenario produces a slightly perturbed metric
that leads to the PPN parameters $ \be \approx 1 - (K + K_r)/ M$
and $\ga \approx 1 - (K + 2 K_r) / M$, directly proportional to
the strength of the derived effect (given by $K$ and $K_r \sim
K$). Since $K$ and $K_r$ are integration constants, no constraints
on the physical parameters may be derived from the observed limits
on the PPN parameters. Also, an analogy with Rosen's bimetric
theory, yields the PPN parameter $\ga \simeq (A + B)\xi$, $\xi$
being the distance to the central body and $A$ and $B$ parameters
ruling the temporal and radial components of the vector field
\textit{vev}.

In the temporal/axial scenario, a breakdown of isotropy is
obtained, disallowing a standard PPN analysis. However, the
direction-dependent ``PPN'' parameters $ \ga_1 \simeq b^2 cos^2
\th /2$ and $\ga_2 \simeq a^2 b^2 cos^2 \th / 4$ may be derived,
where $a$ and $b$ are respectively the temporal and $x$-component
of the bumblebee vector \textit{vev}; naturally, $\ga_1 \ll
\ga_2$. A crude estimative of the PPN parameter $\ga$ yields $ \ga
\approx b^2(1-e^2)/4$, where $e$ is the orbit's eccentricity.
Furthermore, a comparison with experiments concerning the
anisotropy of inertia produces the bounds $ |b| \leq 2.4 \times
10^{-11} $.


\begin{thebibliography}{99}

\bibitem{Kostelecky1} ``CPT and Lorentz Symmetry II'', Ed., V. A. Kosteleck\'{y} (World
Scientific, Singapore, 2002).

\bibitem{Bertolami} O. Bertolami, Nucl.\ Phys.\ Proc.\ Suppl.\ {\bf 88}, 49
(2000); O. Bertolami in ``Decoherence and Entropy in Complex
Systems'' (Springler-Verlag, Berlin, 2004).

\bibitem{Kostelecky2} V. A. Kosteleck\'{y} and S. Samuel, \PR {\bf D 39}, 683
(1989); \PRL {\bf 66}, 1811 (1991).

\bibitem{Kostelecky3} V. A. Kosteleck\'{y} and R. Potting, \PR {\bf D 51}, 3923 (1995).

\bibitem{Kostelecky4} V. A. Kosteleck\'{y}, \PR {\bf D 69}, 105009 (2004).

\bibitem{Kostelecky5} R. Bluhm and V. A. Kosteleck\'{y}, hep-th/0412320.

\bibitem{LSB4} S. M. Carroll, J. A. Harvey, V. A. Kosteleck\'{y}, C. D.
Lane and T. Okamoto, \PRL {\bf 87}, 141601 (2001).

\bibitem{LSB5} O. Bertolami and L. Guisado, \PR {\bf D 67}, 025001 (2003); \JHEP {\bf 0312},
O13 (2003); O. Bertolami, \MPL {\bf A 20}, 1359 (2005).

\bibitem{LSB6} V.A. Kosteleck\'{y}, R. Lehnert and M. J. Perry, \PR {\bf D 68}, 123511
(2003) ; O. Bertolami, R. Lehnert, R. Potting and A. Ribeiro, \PR
{ \bf D 69}, 083513 (2004).

\bibitem{LSB7} H. Sato and T. Tati, \PTP {\bf 47}, 1788 (1972); S.
Coleman and S.L. Glashow, \PL {\bf B 405}, 249 (1997); \PR {\bf D
59}, 116008 (1999); O. Bertolami and C.S. Carvalho, \PR {\bf D
61}, 103002 (2000); O. Bertolami, \GRG {\bf 34} 707 (2002); R.
Lehnert, hep-ph/0312093.

\bibitem{paper} O. Bertolami and J. P\'aramos, \PR {\bf D 72}, 044001 (2005).

\bibitem{Bepi} R. Grard, M. Novara and G. Scoon, \textit{ESA Bull.} {\bf 103}, 11
(2000); L. Iorio, I. Ciufolini and E. C. Pavlis, \CQG {\bf 19},
4301 (2002).

\bibitem{LATOR} S. G. Turyshev \textit{et al.}, gr-qc/0505064.

\bibitem{paper2} O. Bertolami, J. P\'aramos and S. G. Turyshev,
gr-qc/0601016.

\bibitem{Bento} M. C. Bento and O. Bertolami, \PL {\bf B 228}, 348
(1999).

\bibitem{Fischbach} E.
Fischbach and C.L. Talmadge, ``The search for non-Newtonian
gravity'' (Springer, New York 1999).

\bibitem{Will2} C. M. Will, Living Rev.\ Rel.\ {\bf 4}, 4 (2001).

\bibitem{Bertotti} B. Bertotti, L. Iess and P. Tortora, Nature {\bf 425}, 374 (2003).

\bibitem{Rosen} N. Rosen, \textit{J. Gen. Rel. and Grav.} {\bf 4}, 435 (1973).

\bibitem{Will} C.M. Will, ``Theory and Experiment in Gravitational Physics'',
C.M. Will (Cambridge U. P., 1993).

\bibitem{Weinberg} S. Weinberg, ``Gravitation and Cosmology: Principles and Applications of the
General Theory of Relativity'' (John Wiley and Sons, New Jersey,
1972).

\bibitem{Kostelecky6} V. A. Kosteleck\'y and C. D. Lane, J. Math. Phys. {\bf 40}
6245  (1999).

\bibitem{Lamoreaux} S. K. Lamoreaux, J. P. Jacobs, B. R. Heckel, F. J. Raab, and E. N. Fortson,
\PRL {\bf 58}, 746 (1987).


\end{thebibliography}
\end{document}